\documentclass[11pt]{article}
\usepackage{amssymb}

\usepackage{graphicx}

\begin{document}

\author{X. G. Wen$^{1,3}$ \& A. Zee$^{2,3}$ \\
$^{1}$ Department of Physics, Massachusetts Institute of Technology\\
Cambridge, Massachusetts 02139, USA \\
$^{2}$Institute for Theoretical Physics, University of California\\
Santa Barbara, California 93106, USA\\
$^{3}$ Center for Advanced Study, Tsinghua University\\
Beijing 100084, China}
\title{Superfluidity and Superconductivity in Double-Layered Quantum 
Hall state}
\date{}
\maketitle

\begin{abstract}
We discuss and review the basic physics that leads to
superfluidity/superconductivity in certain quantum Hall states, in
particular the so-called double-layered $(mmm)$ state. In the $K$-matrix
description of the quantum correlation in quantum Hall states, those states
with det$(K)=0$ contain a special correlation that leads to
superfluidity/superconductivity. We propose a four-terminal measurement to
test the DC Josephson-like effect in interlayer tunneling, so that the issue
of superfluidity/superconductivity in the $(mmm)$ states can be settled
experimentally.
\end{abstract}

Almost 10 years ago we predicted\cite{wz1,wz2,wz3,wz4} that under certain
circumstances a double layered quantum Hall system would exhibit the physics
of superfluidity/superconductivity. Recently, this apparently surprising and
somewhat counter-intuitive prediction was verified dramatically by
Eisenstein et al\cite{eisen} in a form of (two-terminal) DC Josephson-like
effect for interlayer tunneling.

Historically, the gapless mode in the $(111)$ double layered quantum Hall
state was first discovered by Fertig\cite{Fertig}. Later, the effect of
interlayer tunneling was included\cite{MPB,B}. However, 
superfluidity and superconductivity 
were not discussed in these earlier work. In Ref. \cite{wz1,wz2}, we
discovered and identified the spontaneously broken $U(1)$ symmetry in the $%
(mmm)$ states and emphasized the resulting superfluidity and
superconductivity. The properties we discussed include the
Kosterlitz-Thouless transition, the zero resistivity in the counter-current
flow, the DC/AC Josephson-like effect in interlayer tunneling, as well as
the gapless superfluid mode. Some of these results were obtained later using
an isospin approach for the special $m=1$ case\cite{pspin}.

We stressed that the Josephson-like effect we predicted in the $(mmm)$ state
differ in some crucial respects from the Josephson effect for superconductor
junction\cite{wz2}. The critical current for the $(mmm)$ state is
proportional to the electron tunneling amplitude, rather than the square of
the tunneling amplitude as in the DC Josephson effect between
superconductors. More strikingly, the Josephson frequency in the AC
Josephson-like effect for the $(mmm)$ states is given by \cite{wz2}
\begin{equation}
f_{J}^{*}=eV/h,
\end{equation}
half of the standard Josephson frequency $f_{J}=2eV/h$ in the AC Josephson
effect between superconductors.

There appears to be some confusion and disagreement in the literature
concerning the superfluidity/superconductivity in the $(mmm)$ states. It has
been argued in Ref. \cite{MZ} that ``although gapless out-of-phase modes
associated with a spontaneously broken continuous symmetry can occur in
double-layer systems they do not imply superfluid behavior''. In particular,
the DC Josephson-like effect in the interlayer tunneling remain a
controversial issue\cite{MZ,JM}. Since the physics underlying the
superfluidity/superconductivity in double-layered quantum Hall system is
both fundamental and profound, we would like to clarify and review some of
the basic principles. Some of what we say here can be found in one form or
another in our original paper and in various reviews\cite{wentopo,zrev} we
wrote. Still, we hope that our presentation here will help to shed light on
the basic physics.

In this paper, we will propose a four-terminal set-up to test the DC
Josephson-like effect experimentally. We hope that future experiments will
clarify the controversy on the DC Josephson-like effect and the
superfluidity/superconductivity in the $(mmm)$ states.

For conceptual clarity, our discussion here will be entirely at zero
temperature.

\section{A dual description of the superfluid}

As was emphasized by Feynman\cite{feyn} among others, the physics of
superfluidity lies not in the presence of gapless excitations, but in the
paucity of gapless excitations. After all, the Fermi liquid has a continuum
of gapless modes. Indeed, consider a gas of free bosons at zero temperature.
We can give a momentum $\hslash \vec{k}$ to any given boson at the cost of
only $\hslash \vec{k}^{2}/2m$ in energy. There exist many low energy
excitations in a free boson system. But as soon as a short-ranged repulsion
is turned on between the bosons, a boson moving with momentum $\vec{k}$
would affect all the other bosons. A density wave is set up as a result. As
Bogoliubov and others have taught us, the density wave has energy $%
\varpropto k$. The gapless mode has gone from quadratically dispersing to
linearly dispersing. There are far fewer low energy excitations.
Specifically, the density of states $N(E)\varpropto k^{D-1}(\frac{dk}{dE})$
goes from $N(E)\varpropto $ constant to $N(E)\varpropto E$ at low energies.
Here we have set the dimension of space $D$ to 2.

If we accept the point of view that the key ingredient of superfluidity is
the existence of \textit{only} one mode of gapless excitations, then we can
use this picture to give a semi-quantitative description of superfluid. As
the above discussion makes clear, the linearly dispersing gapless mode is a
density wave. As Landau first proposed, such a density fluctuation can be
described hydrodynamically. The kinetic energy density is then simply given
by the square of the current density $\vec{j}$, while the potential energy
density is given by $(\delta \rho )^{2}$ where $\delta \rho $ represents the
deviation of number density $\rho $ from some mean density $\rho _{0}$.
Thus, suppressing some constant factors (which we can set to unity by a
suitable choice of spacetime units), we have the effective Lagrangian for
the superfluid $\mathcal{L}=\vec{j}^{2}-(\delta \rho )^{2}+\cdots $ where as
always in an effective Lagrangian description the $(\cdots )$ represent
physics that are not important at low energy and momentum. Suppressing the $%
\delta $ on $\delta \rho $ and using the relativistic notation $j^{\mu
}=(\rho ,\vec{j})$ we can write
\begin{equation}
\mathcal{L}=-j^{\mu }j_{\mu }+\cdots  \label{hydro}
\end{equation}

A crucial physical principle underlying a hydrodynamic description is of
course current conservation $\frac{\partial \rho }{\partial t}-\vec{
\triangledown}\vec{j}=0,$ or in relativistic notation
\begin{equation}
\partial _\mu j^\mu =0.
\end{equation}
This conservation equation is solved by writing
\begin{equation}
j^\mu =\frac 1{2\pi }\varepsilon ^{\mu \nu \lambda }\partial _\nu a_\lambda
\label{current}
\end{equation}
We recognize the symbol $a_\lambda $ as representing a gauge potential since
the transformation
\begin{equation}
a_\lambda \rightarrow a_\lambda +\partial _\lambda \Lambda
\end{equation}
leaves the physical current $j^\mu $ and hence physics unchanged. The
relation (\ref{current}) is an example of a duality relation.

Upon substituting (\ref{current}) into (\ref{hydro}) we obtain the effective
Lagrangian in terms of the gauge potential $a_\lambda$
\begin{equation}
\mathcal{L}_{superfluid}=-\frac 1{8\pi ^2}f^{\mu \nu }f_{\mu \nu }+\cdots
\label{max}
\end{equation}
where $f_{\mu \nu }=\partial _\mu a_\nu -\partial _\nu a_\mu$. We refer to (%
\ref{max}) as the Maxwell Lagrangian since it has the same form as the
Lagrangian introduced by Maxwell to describe electromagnetism.

We could have described the gapless linearly dispersing mode by the equation
of motion $\partial _{\mu }f^{\mu \nu }=0$ and thus gone to (\ref{max})
directly. However, it would have been unclear why we would need a spin 1
field to describe one gapless mode, even though it is true that in $(2+1)$
dimensional spacetime a spin $1$ field contains only one physical degree of
freedom. This gapless mode is in fact a Nambu-Goldstone boson. Since the
only locally conserved quantity in our interacting Bose gas is the number
density $\rho $ associated with a global $U(1)$ symmetry, this global
symmetry must be spontaneously broken. This is indeed so as the number of
bosons in the $\vec{k}=0$ ground state is a finite fraction of the total
number of bosons.

The physics of gapless mode is particularly transparent in Bogoliubov's
calculation. Let $a_{0}$ and $a_{0}^{\dagger }$ be the annihilation and
creation operators for bosons in the $\vec{k}=0$ non-interacting ground
state. In the interaction term $\sum\limits_{k,p,q}G(a_{k}^{\dagger
}a_{p}^{\dagger }a_{q}a_{k+p-q})$ in the Hamiltonian $H$ Bogoliubov replaces
$a_{0}$ and $a_{0}^{\dagger }$  by $c-$numbers $<a_{0}>=<a_{0}^{\dagger }>=u,
$ thus giving schematically
\begin{equation}
H\sim \sum_{k\neq 0}Gu^{2}(a_{k}^{\dagger }a_{k}+a_{-k}^{\dagger
}a_{-k}+a_{k}^{\dagger }a_{-k}^{\dagger }+a_{-k}a_{k})+\sum_{k}\frac{k^{2}}{%
2m}a_{k}^{\dagger }a_{k}
\end{equation}
Clearly, we get a gapless mode since up to an additive constant the first
term can be written as $\sum_{k\neq 0}Gu^{2}(a_{k}^{\dagger
}+a_{-k})(a_{k}+a_{-k}^{\dagger })$ and so $a_{k}^{\dagger }-a_{-k}$ creates
a gapless excitation of momentum $\vec{k}.$ The crucial point is that  $%
<a_{0}>=<a_{0}^{\dagger }>\neq 0$ and not merely $<a_{0}^{\dagger }a_{0}>%
\neq 0$ : in that case we would only get a gapped mode with energy $(2Gu^{2}+%
\frac{k^{2}}{2m})$.

Formally, the $U(1)$ symmetry breaking can be described by a field theoretic
representation of the boson gas
\begin{equation}
\mathcal{L}=\partial \Phi ^{\dagger }\partial \Phi -V(\Phi ^{\dagger }\Phi )
\end{equation}
A superfluid is defined as a fluid in which $<\Phi >=we^{i\theta }$ does not
vanish. As is well known, this breaks the global $U(1)$ symmetry $\Phi
\rightarrow e^{i\alpha }\Phi $ of the Lagrangian $\mathcal{L}$. %
%
In this language, it is possible to describe the gapless linearly dispersing
mode by a scalar field and to write 
\begin{equation}
\mathcal{L}_{superfluid}=\frac{c}{2}(\partial _{\mu }\theta )^{2}+\cdots
\label{theta}
\end{equation}
where the phase field is the phase of the field $\Phi $.
We note that the Lagrangian (\ref{max}) and (\ref{theta}) describe the same
superfluid mode and the two theories are dual to each other.

%

\section{Quantum Hall Fluid}

The Hall fluid is another highly coherent quantum state. To see how the low
energy effective theories of the Hall fluid and of the superfluid differ
from each other, we need to understand the correlations in the quantum Hall
ground state. The simplest kind of Hall fluid with filling fraction $\nu
=1/k $ is described by the ground state wave function proposed by Laughlin
\begin{equation}
\Psi \sim \prod\limits_{i,j}(z_{i}-z_{j})^{k}  \label{lau}
\end{equation}
with $k$ an odd integer. This highly non-trivial wave function captures the
strong quantum correlation between particles. As electron $i$ goes around
electron $j$ the wave function acquires a factor $e^{2i\pi k}.$ As is well
known but still worth emphasizing, the essential physics involved is purely
quantum in character and has no classical counterpart.

This quantum phase correlation is the same as in the Aharonov-Bohm effect
and so the particles in (\ref{lau}) can be regarded as carrying both charge
and flux associated with some gauge field. As was shown by Wilczek and Zee%
\cite{wilzee}, this phase correlation can be expressed mathematically as a
Hopf term which in turn can be generated by a Chern-Simons term in the gauge
action. The phase correlation can also be viewed as a density-current
interaction.

The origin of the Hopf and Chern-Simon terms can be easily understood in
physical terms within the context of this discussion. When electron $i$ goes
around electron $j$, we have a current $\vec{j}$ around a charge density $%
\rho $ (as before, we drop the $\delta $), and so the action $S$ should
contain a non-local term of the form $\sim $ $\int d^{3}x$ $\int
d^{3}x^{\prime }$ $\rho (x)\vec{K}(x-x^{\prime })\vec{j}(x^{\prime })$ in
addition to the terms in (\ref{hydro}). Such a term describes a non-local
current-density interaction. The kernel $\vec{K}$ can be determined from
basic principles. Rotational invariance and scale invariance fix this term
to be $S_{Hopf}$ $\sim $ $\int d^{3}x$ $\int d^{3}x^{\prime }$ $k\rho (x)%
\frac{1}{|x-x^{\prime }|}\vec{\triangledown}\times \vec{j}(x^{\prime })$ or
in relativistic form 
\begin{equation}
S_{Hopf}\sim k\int d^{3}x\varepsilon ^{\mu \nu
\lambda }j_{\mu }\frac{1}{\partial ^{2}}\partial _{\nu }j_{\lambda }.
\end{equation}
We have scale invariance because we know that the phase factor $e^{i(2\pi m)}$
does not depend on the separation between particles $i$ and $j$: there is no
characteristic distance in the quantum phase correlation. Note that the
cross product $\vec{\triangledown}\times \vec{j}$ is forced on us by
rotational invariance and current conservation: if instead we wrote $\vec{%
\triangledown}\cdot \vec{j}$ we would have a density density interaction
instead. It follows immediately that $S_{Hopf}$ violates time reversal.

The non-locality of the Hopf action can be removed by introducing a gauge
potential $a^{\mu }$ to represent $j^{\mu }$ (see (\ref{current})).
After substituting $j^{\mu}$ with $a^\mu$, the Hopf term becomes the
Chern-Simons term:
\begin{equation}
S_{Chern-Simons}=\int d^{3}x\ \frac{k}{4\pi }
\varepsilon ^{\mu \nu \lambda }a_{\mu }\partial _{\nu }a_{\lambda
} 
\end{equation}
Thus, the effective action for the $1/k$ Laughlin state is
\begin{equation}
\mathcal{L}_{Hall}=\frac{k}{4\pi }\varepsilon ^{\mu \nu \lambda }a_{\mu }%
\partial _{\nu }a_{\lambda }-\frac{1}{4g^{2}}f^{\mu \nu }f_{\mu \nu }+\cdots
\label{hall}
\end{equation}
where we have restored the coefficient of the Maxwell term (which we should
strictly write in non-relativistic form as $%
(f_{0i})^{2}-v^{2}(f_{ij})^{2}=e^{2}-v^{2}b^{2}$ with some characteristic
velocity $v$). This effective action\cite{wentopo,zrev} reproduces the
correct Hall conductance $\sigma =\frac{e^{2}}{2\pi k}$ if we include the
coupling to the electromagnetic gauge potential $A_{\mu }$ by adding $%
eA_{\mu }j^{\mu }=\frac{e}{2\pi }\varepsilon ^{\mu \nu \lambda }A_{\mu }%
\partial _{\nu }a_{\lambda }$ to $\mathcal{L}$ .

The quantum correlation (or the density-current interaction) is not a
potential energy between particles and thus cannot be expressed in energetic
terms. This physical observation is manifested mathematically by the fact
that the Chern-Simons term contains only the Levi-Civita symbol $\varepsilon
^{\mu \nu \lambda }$ and does not contain the metric $g^{\mu \nu }$ and so
according to Einstein the corresponding term in the Hamiltonian is
identically zero.

Since the Chern-Simons term contains only one spacetime derivative $\partial %
_{\nu }$ while the Maxwell term $f^{\mu \nu }f_{\mu \nu }$ contains two, the
Chern-Simons term completely dominates the Maxwell term at low energy and
momentum. The effective theory of $1/k$ Laughlin state can be written as $%
\mathcal{L}_{Hall}=\frac{k}{4\pi }\varepsilon ^{\mu \nu \lambda }a_{\mu }%
\partial _{\nu }a_{\lambda }+\cdots $

Counting powers of derivatives in $\mathcal{L}_{Hall}$, we see that the
inverse propagator for a gauge boson of energy-momentum $q$ has the
schematic form (we suppress Lorentz indices and display only the power of
momentum) $\sim kq-\frac{1}{g^{2}}q^{2}$ and thus the propagator has a pole
at $q\sim kg^{2}.$ The gauge boson has a mass $\sim kg^{2}$ or in
non-relativistic terms the spectrum of the Hall fluid has a gap $\sim kg^{2}$%
. In the limit $g\rightarrow \infty ,$ the spectrum contains only
zero-energy ground states. As is well known by now, $\mathcal{L}%
_{Chern-Simons}=\frac{k}{4\pi }\varepsilon ^{\mu \nu \lambda }a_{\mu }%
\partial _{\nu }a_{\lambda }$ is an example of a topological field theory.
The only physical question we can ask of the spectrum is how many ground
states there are given appropriate boundary conditions. With various spatial
boundary conditions, the theory is effectively compactified on Riemann
surfaces of genus $g$ (not to be confused with the Maxwell coupling of
course.) The question is then the topological ground state 
degeneracy\cite{wentopo} $D$ as a function of the genus $g.$

Contrast $\mathcal{L}_{Hall}$ with $\mathcal{L}_{superfluid}$. The presence
of the Chern-Simons term (or equivalently, the density-current interaction)
completely overwhelms the Maxwell term. There is no gapless mode in the
bulk. The Hall fluid is incompressible and the only gapless modes must
reside on the edge\cite{Wedge}.

\section{Double Layered Hall Fluid}

Given the above discussion it would seem surprising that quantum Hall
systems can under some circumstances contain a superfluid. We will now
explain how this could happen in a double layered system. Conceptually, it
is clearer to start with no (or better, an infinitesimal) interlayer
tunneling.

In a double layered system we can associate a separate gauge potential $%
a_{\mu }^{I}$ ($I=1,2)$ with each layer. This is because in the absence of
tunneling the current in each layer is separately conserved and thus we can
write (\ref{current}) separately for each layer: $j^{I}_{\mu }=\frac{1}{2\pi }%
\varepsilon ^{\mu \nu \lambda }\partial _{\nu }a_{\lambda }^{I}$ $(I=1,2).$
Thus, the effective Lagrangian (\ref{hall}) is naturally generalized to
\begin{equation}
\mathcal{L}_{Hall}=\sum_{I,J}(\frac{K_{IJ}}{4\pi }\varepsilon ^{\mu \nu
\lambda }a_{\mu }^{I}\partial _{\nu }a_{\lambda }^{J}-\frac{1}{4}%
(g^{-2})_{IJ}f^{I}_{\mu \nu }f_{\mu \nu }^{J}+\cdots )
\end{equation}
We have merely put layer indices on the various fields. Significantly,
however, the number $k$ has been promoted to a matrix $K$. Over the years,
we and others\cite{Kmatrix} have studied this $K$-matrix description of
quantum Hall fluids. In particular, we find that all Abelian quantum Hall
fluids can be classified by the $K$-matrices\cite{clss}.

With the matrix
\[
K=\left(
\begin{array}{ll}
k & m \\
m & l
\end{array}
\right)
\]
$\mathcal{L}_{Hall}$ describes the double-layered $(klm)$ state with the
wave function
\begin{equation}
\Psi \sim
\prod\limits_{i,j}(z_{i}-z_{j})^{k}\prod\limits_{i,j}(w_{i}-w_{j})^{l}\prod%
\limits_{i,j}(z_{i}-w_{j})^{m}
\end{equation}
where $z_{j}$ and $w_{j}$ denote the coordinates of the electrons in layer 1
and 2 respectively. Note that the factor $\prod%
\limits_{i,j}(z_{i}-w_{j})^{m} $ describes a subtle quantum correlation
between the two layers.

In general the effective theory $\mathcal{L}_{Hall}$ describes a quantum
Hall state with finite energy gap. But when $k=l=m$, the matrix $K$ can have
a zero eigenvalue! The corresponding eigenvector now describes a linear
combination of gauge potentials, $a_{-,\mu }=a_{\mu }^{1}-a_{\mu }^{2}$,
which is no longer governed at low energy and momentum by the Chern-Simons
term, but by the Maxwell term. Thus the low energy effective theory for the $%
(mmm)$ quantum Hall state is identical to the low energy effective theory of
a superfluid (\ref{max}). The gauge potential $a_{-,\mu }$ corresponds to a
gapless linearly dispersing mode. Thus, the double layered $(mmm)$ quantum
Hall state is in fact a superfluid state.

Conceptually, this is associated with the appearance of an off-diagonal long
range order cause by a spontaneously broken $U(1)$ symmetry. The broken
symmetry can be identified by noting that $a_{-,\mu }$ describe the
fluctuations of the difference of the electron numbers $N_{1}-N_{2}$ in the
two layers. Thus it is the symmetry associated with the conservation of $
N_{1}-N_{2}$ that is broken in the $(mmm)$ state.

In the broken symmetry phase, the order parameter $<c_{1}^{\dag
}c_{2}>\propto e^{i\theta }$ is non-zero even in the absence of interlayer
tunneling (here $c_{1}$ and $c_{2}$ denote the electron annihilation
operator in layer 1 and 2 respectively). The fluctuations of the angle field
$\theta $ give rise to a gapless mode described by the standard X-Y model
for any $U(1)$ symmetry breaking phase:
\begin{equation}
\mathcal{L}=\frac{c}{2}((\partial _{t}\theta )^{2}-v^{2}(\partial _{i}\theta
)^{2})  \label{XY}
\end{equation}
Here $c$ may be thought of as the capacitance per unit area between the two
layers.

The $U(1)$ gauge description allows us to conclude that any quantum Hall
state with det$K=0$ may have a spontaneously broken $U(1)$ symmetry and
exhibit superfluidity/superconductivity. This result is much more general
than the one obtained from the isospin approach\cite{pspin}. The X-Y model
description is more convenient for studying interlayer tunneling.

When interlayer tunneling is ``switched'' on, the $U(1)$ symmetry associated
with $N_{1}-N_{2}$ is explicitly broken. 
In the limit of weak tunneling amplitude,
we obtain the effective Lagrangian by including the tunneling term $%
<c_{1}^{\dagger }c_{2}+h.c.>=\eta \cos \theta $:
\begin{equation}
\mathcal{L}=\frac{c}{2}((\partial _{t}\theta )^{2}-v^{2}(\partial _{i}\theta
)^{2})+\eta \cos \theta .  \label{w2}
\end{equation}
Various physical quantities can be expressed in terms of the $\theta $ field%
\cite{wz1,wz2}. The difference in electron density $n_{-}=n_{1}-n_{2}$ and
the difference in electron current density $j_{-}$ between the two layers
are given by
\begin{equation}
n_{-}=2c\partial _{t}\theta ,\ \ \ \ \ j_{-,i}=2v^{2}c\partial _{i}\theta
\label{w3}
\end{equation}
The tunneling current density between the two layers is given by
\begin{equation}
j_{T}=\eta \sin \theta   \label{w4}
\end{equation}
The voltage difference between the two layers is
\begin{equation}
V=\frac{\hbar \partial _{t}\theta }{e}  \label{w5}
\end{equation}

We see that a static solution $\theta \neq 0$ corresponds to a finite
tunneling current $j_{T}=\eta \sin \theta $ even for vanishing voltage
difference between the two layers. This is reminiscent of the DC Josephson
effect in the tunneling between two superconductors, but as noted in 
Ref. \cite{wz4} there is a crucial difference. In the case of two
superconductors, we have instead of (\ref{w2})
\begin{equation}
\mathcal{L}=\sum_{J=1,2}\frac{c_{J}}{2}((\partial _{t}\theta
_{J})^{2}-v_{J}^{2}(\partial _{i}\theta _{J})^{2})+\eta \cos (\theta
_{1}-\theta _{2})
\end{equation}
where $c_{J}$ scales with the volume of the superconductors and $\eta $ with
the area of the contact between the superconductors. In contrast, in (\ref
{w2}) $c$ and $\eta $ both scale with the area of the quantum Hall system.
Thus, in the double layered quantum Hall system interlayer tunneling opens
up a finite energy gap $\sim \sqrt{\frac{\eta }{c}}.$ In the case of two
superconductors, the would-be energy gap scales to $0$ in the limit of large
system size.

\section{Four terminal measurement -- experimental test of DC Josephson-like
effect}

In this section we would like to consider how to experimentally test the
prediction that a small interlayer tunneling current does not induce any
voltage drop between the two layers for the $(mmm)$ state at $T=0$. This
test is important, since the DC Josephson-like effect we predicted in \cite
{wz1, wz2, wz3, wz4} is still a controversial issue. It was argued in Ref.
\cite{MZ,JM} that the $(111)$ state does not have any Josephson effect in
the presence of interlayer tunneling.

\begin{figure}[tbp]
\centerline{\ \includegraphics[width=3in]{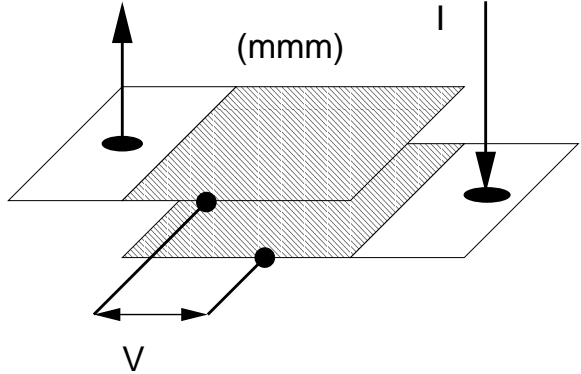} }
\caption{Four terminal set-up for testing DC Josephson-like effect}
\label{fig1}
\end{figure}

One way to test the zero voltage drop is through a sample with the geometry
described in Fig. \ref{fig1}. The middle region is occupied by the $(mmm)$
state, and the left and the right region by a metallic state (\textit{e.g.}
the $\nu =1/2$ state) in the first and second layer respectively. The
metallic state in the R and L region serve as ohmic contacts to the first
and second layer in the $(mmm)$ state. One voltage probe is attached to an
edge of the $(mmm)$ state in the first layer, and the other attached to the
edge on the same side in the second layer (see Fig. \ref{fig1}). As we pass
a current from the right region to the left region, the current must tunnel
through the barrier between the two layer. According to Ref. \cite{wz2}, if
the current is less than a critical value (here we assume a uniform
interlayer tunneling), the system is described by a time-independent $\theta
$ in (\ref{w2}). From (\ref{w5}), we see that such a static configuration
does not induce any voltage between the two layers. In other words, the
voltage drop between two probes in the two layers would be zero if the
probes are on top of each other (\textit{i.e.} have the same $x$-$y$
coordinates). However, such probes are hard to make in experiments. Since
the interlayer tunneling induces a finite energy gap in the bulk of the
sample, the chiral edge state does not have any voltage drop along a single
edge. Thus we can place the two probes at different place along the edge, as
shown in Fig. \ref{fig1}. The voltage drop between the two voltage probes
should be zero. For a larger current beyond a critical value, a non-zero voltage
drop will be developed. The striking feature is that the voltage drop is zero even
for weak interlayer tunneling. Weak interlayer tunneling only leads to a
small critical current.

We would like to make four remarks. \newline
(1) If the second voltage probe is attached to the edge on the other side of
the sample from the first voltage probe, a non-zero voltage drop will appear
due to the non-zero Hall resistance. \newline
(2) The vanishing of the voltage drop between the two leads is a consequence
of the superfluidity , not of the chirality of the edge excitations (%
\textit{i.e.} the absence of back propagating modes). To illustrate this
point let us replace the $(mmm)$ state in the middle region by a $(mm0)$
state, which does not have any superfluidity (\textit{i.e.} its $K$-matrix
does not have a zero eigenvalue). There is no quantum coherence between the
two separate $1/m$ states in the two layers. Now interlayer tunneling is
controlled by tunneling between the two edges of the two $1/m$ states. In
this case a finite current will produce a finite voltage drop between the
two layers and hence between the two voltage probes. If the edges are sharp
on the scale of the magnetic length, we have\cite{Wedge} $V\propto
I^{1/(2m-1)}$ at zero temperature $T=0$ and $(\frac{dI}{dV})_{V=0}\propto
T^{2m-2}$.\newline
(3) In the strong tunneling limit, the $(mmm)$ state crosses over to a
single layered $\nu =1/m$ state. The zero voltage drop between the two leads
is just the standard result $\rho _{xx}=0$ for fractional quantum Hall
states. This correct cross-over behavior from the weak to the strong
tunneling limit further confirms the validity of our results.\newline
(4) If the interlayer tunneling is not uniform, the flux trapped between the
layers should be pinned to observe the DC Josephson-like effect\cite{BR,FW}.

Before closing this section, we would like to mention that in a two-terminal
measurement as arranged in Fig. 1, the measured two-terminal resistance should
be exactly $mh/e^2$ for the $(mmm)$ state if the current is smaller than a
critical value.  Note that the small-current two-terminal resistance is
always $mh/e^2$ regardless the strength the interlayer tunneling amplitude,
despite the tunneling current flows across the interlayer junction. This
resembles the zero DC resistance of a Josephson junction between two
superconductors at small currents.  A small
interlayer tunneling only reduces the value of the critical current. For
current beyond the critical value, the two-terminal resistance will become very
large in the weak tunneling limit.

\section{Quantized Hall drag resistance}

Recently Kellogg {\it et al}\cite{Kellogg} has observed a quantized Hall drag
resistance in $(111)$ state, as predicted by several
groups.\cite{Moon,Duan,Yang,Kim} Here we would like to show that quantized
Hall drag resistance can be easily obtained using the $K$-matrix Chern-Simon
effective theory. This allows us to calculate the Hall drag for any Abelian
quantum Hall state characterized by $K$.

For simplicity, let us consider a double layer state characterized by a
$2\times 2$ $K$-matrix. The effective theory is given by
\begin{equation}
\mathcal{L}_{Hall}=\sum_{I,J}(\frac{K_{IJ}}{4\pi }\varepsilon ^{\mu \nu
\lambda }a_{\mu }^{I}\partial _{\nu }a_{\lambda }^{J})+
\sum_{I}(\frac{1}{2\pi }\varepsilon ^{\mu \nu
\lambda }a_{\mu }^{I}\partial _{\nu }A_{\lambda }^{I})
\end{equation}
where $A^I_{\mu}$, $I=1,2$, is the electromagnetic gauge potential in the
$I^{th}$ layer. From the equation of motion, we find that the induced current
in the two layers, $j^{I}_{\mu} = \frac{1}{2\pi } \varepsilon ^{\mu \nu
\lambda }\partial _{\nu }a_{\lambda }^{I}$, are given by
\begin{equation}
 j^I_{i} = \frac{e^2}{h} (K^{-1})^{IJ}
\varepsilon ^{ij}E_j^J
\end{equation}
where $E_i^I$ is the electric field in the $I^{th}$ layer, and $i,j=x,y$.  We
can define the Hall drag resistance, $R_{xy,IJ}$, as the coefficiant that
determine the Hall voltage in the $I^{th}$ layer induced by the current in the
$J^{th}$ layer
\begin{equation}
 E_i^I = R_{xy,IJ} \varepsilon^{ij} j_i^J
\end{equation}
We find $R_{xy,IJ}$ is given by
\begin{equation}
 R_{xy,IJ} = \frac{h}{e^2} K_{IJ}
\end{equation}
The above result applies to any double-layer quantum Hall state characterized
by the $2\times 2$ $K$-matrix.  It is interesting to see that the $K$-matrix
can be directly measured via the Hall drag matrix.  For the $(klm)$ state,
$K_{11} = k$, and $K_{12}=m$. We have
\begin{equation}
 R_{xy,11} = \frac{kh}{e^2}, \ \ \ \
 R_{xy,12}=  \frac{mh}{e^2}
\end{equation}
which is exactly what was observed by Kellogg {\it et al} for the $(111)$ state.

\section{Summary}

The superfluidity/superconductivity in certain quantum Hall state reveals a
subtle quantum correlation. Using the $K$-matrix description of the quantum
Hall states, we can identify exactly which quantum Hall states have such
special correlation. We find that if $K$ has a zero eigenvalue, then the
corresponding quantum Hall state may have a spontaneously broken $U(1)$
symmetry and the associated superfluidity/superconductivity. The picture
presented in this paper reveal the basic physics behind the
superfluidity/superconductivity in quantum Hall states. Our formalism can be
applied to more general quantum Hall systems well beyond the double layer
system, to multi-layered systems for example. We also proposed a concrete
four-terminal test for the DC Josephson-like effect. We hope that the
controversy around the superfluidity/superconductivity in the $(mmm)$ state
and the DC Josephson-like effect in the interlayer tunneling can be settled
by experiments. Also we would like to point out that observing the AC
Josephson-like effect (a peak at $f_{J}^{*}$ in the noise spectrum of
interlayer tunneling current) will truly reveal the unusual correlation in
the $(mmm)$ state.

\section{Acknowledgment}

We thank Professor H. T. Nieh for his hospitality at the Center for Advanced
Study, Tsinghua University, Beijing, China. XGW is supported by NSF Grant
No. DMR--01--23156 and by NSF-MRSEC Grant No. DMR--98--08941, and AZ by NSF
Grant No. PHY-99-07949.

\end{document}